\documentclass[twocolumn,showpacs,superscriptaddress,amssymb,10pt,prl,floatfix]{revtex4}

\usepackage{graphicx}
\usepackage{dcolumn}
\usepackage{bm}
\usepackage{epsfig}
\usepackage{color}
\usepackage{amsfonts}
\usepackage{amsmath}
\usepackage{mathtools}
\usepackage{caption}
\usepackage{subfig}
\usepackage{transparent}
\begin{document}

\preprint{}
%
%
%
%
\title{Low-energy tests of Delbrück scattering}
%
%
%
%
\author{J.~Sommerfeldt}
\thanks{These two authors contributed equally}
\affiliation{Physikalisch--Technische Bundesanstalt, D--38116 Braunschweig, Germany}
\affiliation{Technische Universit\"at Braunschweig, D--38106 Braunschweig, Germany}

\author{S.~Strnat}
\thanks{These two authors contributed equally}
\affiliation{Physikalisch--Technische Bundesanstalt, D--38116 Braunschweig, Germany}
\affiliation{Technische Universit\"at Braunschweig, D--38106 Braunschweig, Germany}

\author{V.~A.~Yerokhin}
\affiliation{Physikalisch--Technische Bundesanstalt, D--38116 Braunschweig, Germany}
\affiliation{Max-Planck-Institut für Kernphysik, D--69117 Heidelberg, Germany}

\author{W.~Middents}
\affiliation{Helmholtz-Institut Jena, D--07743 Jena, Germany}
\affiliation{Friedrich-Schiller-Universität Jena, D--07737 Jena, Germany}
\affiliation{GSI Helmholtzzentrum für Schwerionenforschung, D--64278 Darmstadt, Germany}

\author{Th.~Stöhlker}
\affiliation{Helmholtz-Institut Jena, D--07743 Jena, Germany}
\affiliation{Friedrich-Schiller-Universität Jena, D--07737 Jena, Germany}
\affiliation{GSI Helmholtzzentrum für Schwerionenforschung, D--64278 Darmstadt, Germany}

\author{A.~Surzhykov}
\affiliation{Physikalisch--Technische Bundesanstalt, D--38116 Braunschweig, Germany}
\affiliation{Technische Universit\"at Braunschweig, D--38106 Braunschweig, Germany}

\date{\today \\[0.3cm]}
%
%
%
%
\begin{abstract}
We present a theoretical study of elastic photon scattering by atomic targets. This process is of special interest since various channels from atomic and nuclear physics as well as quantum elctrodynamics (QED) contribute to it. In this work, we focus on Delbrück scattering which proceeds via production of virtual $e^+e^-$ pairs. In particular, we explore whether and how the Delbrück channel can be ``seen" in present synchrotron experiments which employ strongly linearly polarized light in the energy range of a few hundred keV. In order to answer this question, detailed calculations have been performed for the scattering of 300 keV and 889.2 keV photons off helium-like tin ions. Based on these calculations, we argue that the Delbrück scattering for the energies below the threshold for $e^+e^-$ pair creation leads to a shift in the angular distribution and the polarization of the scattered photons which can be observed by state-of-the-art solid-state detectors.\\
\end{abstract}
%
%
%
%
%
\maketitle

\begin{figure}
\begin{center}
\includegraphics[width=0.85\linewidth]{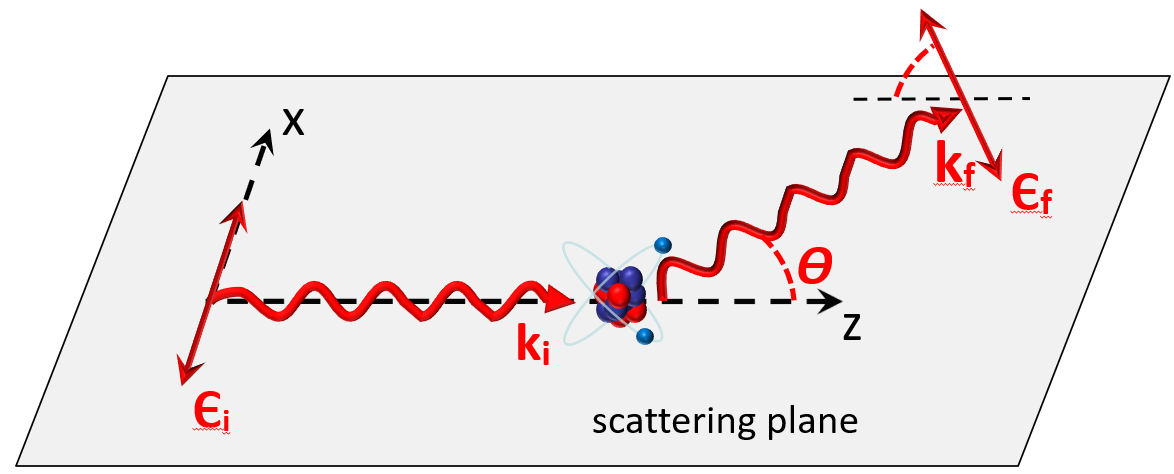}\caption{The geometry of elastic photon scattering by a helium-like target ion. }\label{scatteringplane}
\end{center}
\end{figure}

%
%
\textit{Introduction.}--- The elastic scattering of photons by atomic targets is one of the most fundamental processes in the interaction of light and matter. There are a number of contributions to this process, related to different fields of modern physics. Indeed, the incident photons can be scattered by the bound atomic electrons (Rayleigh, R), by the nucleus (nuclear Thomson, NT) or even by the quantum vacuum via production of virtual electron-positron pairs (Delbrück, D). Moreover, during the recent years, particular attention was paid to the excitation of the giant dipole resonance (GDR) of the nucleus via the scattering. While the Rayleigh and nuclear Thomson channels have been intensively studied in the past \cite{Sc69,BrPe54, Co54, Le51}, much less is known about the Delbrück and GDR contributions. The investigations of these two channels in the low energy regime, i. e. when the incident x-ray energy is below the pair production threshold, is of particular importance. From one side, very little information is available for the low-energy tail of the GDR scattering. From the other side the analysis of Delbrück scattering for the photon energies below 1 MeV is of great interest for probing QED in strong electromagnetic fields. In this energy range is the Delbrück amplitude purely real and thus only the exotic process of vacuum polarization contributes to the scattering cross section.\\
Third generation synchrotron light sources can be used to probe the elastic photon scattering as they provide intense x-ray radiation in well-defined polarization states. These facilities typically provide photon energies up to 500 keV \cite{BiEl05}, which is an additional argument to investigate the low-energy regime of elastic photon scattering. During the last decades the strongly linearly polarized synchrotron radiation has been employed already to study the elastic scattering. In particular, in experiments performed at PETRA III at DESY, both the angular distribution and the polarization of the scattered photons have been measured and provided important information about the electronic structure of the target atoms \cite{BlFr16, MiWe23}. However, there is still an open question whether these measurements can be used to explore the \textit{individual} channels of the elastic scattering.\\
In this letter, we discuss how to uncover the Delbrück contribution to the elastic x-ray scattering in the low-energy regime. Apart from providing an important benchmark for testing non-linear QED in strong fields, the knowledge about the Delbrück channel is critical for extracting the GDR part of the elastic photon scattering \cite{MILSTEIN1994183, RuZu83}. Therefore, we study how the Delbrück contribution may affect the angular distribution and the linear polarization of elastically scattered photons for a typical setup of a scattering experiment at a synchrotron facility. As a target we have chosen a helium-like heavy ion which is the simplest case of a closed-shell many-electron system possessing spherical symmetry. For the helium-like tin ions in particular, we performed calculations for the scattering of 300 keV and 889.2 keV photons by taking into account Rayleigh, nuclear Thomson and Delbrück channels. While the Rayleigh calculations have been performed within the Dirac-Fock screening potential approximation, the Delbrück amplitudes were obtained in all orders of $\alpha Z$ for a pure Coulomb potential. Based on these calculations, we have shown that for the scattering of strongly linearly polarized x-rays the Delbrück channel exhibits itself in the shift of the minima of the angular distribution and the linear polarization which can be detected with state-of-the-art solid-state detectors.\\
Relativistic units (r.u.) $\hbar = m_e = c =1 $ are used throughout this paper, if not
stated otherwise.\\\
%
%
%
%
\textit{Geometry.}--- Before discussing the theory of elastic photon scattering, let us introduce first the geometry of the process. As seen from Fig. \ref{scatteringplane}, the wave vector of the incident photon $\bm k_i$ defines the $z$ axis, and together with the wave vector of the scattered photon $\bm k_f$ spans the scattering ($xz$) plane. For this choice of geometry, the direction of the (outgoing) scattered photons is characterized by the single scattering angle $\theta$. Moreover,  $\bm \epsilon_i$ and $\bm \epsilon_f$ are the polarization vectors of both photons.\\
%
\textit{Theory.}--- As mentioned above, in this work we will consider the elastic x-ray scattering by helium-like ions in their ground state. For such a closed-shell (spherically-symmetric) system, the symmetry considerations suggest that all observables of the scattering process can be expressed in terms of just two amplitudes $A_\parallel$ and $A_\perp$, see Refs. \cite{RoSa86, StYe21}. These amplitudes describe the case when incoming and outgoing photons are both polarized either within or perpendicular to the scattering plane. They can be expressed as the coherent sum of the amplitudes of the contributing scattering channels:

\begin{equation} \label{AmpSum}
    A_{\parallel, \perp}= A_{\parallel, \perp}^{NT}+A_{\parallel, \perp}^R+A_{\parallel, \perp}^D.
\end{equation}
As seen from this expression, we assume that the elastic scattering proceeds via the nuclear Thomson (NT), Rayleigh (R) and Delbrück (D) channels. Giant nuclear resonance will be neglected throughout this work, as its amplitudes and hence the contribution to the cross section are expected to be very small for x-ray energies below the $e^+ e^-$ pair production threshold, considered here \cite{RoSa86, HuLv00}. \\
To evaluate the amplitude \eqref{AmpSum}, we have to discuss the different channels separately. We start with the simplest case of nuclear Thomson scattering whose amplitude can be written as
\begin{subequations}\label{thomson}
\begin{align}
    A^{NT}_\perp &= -\frac{\alpha Z^2}{M}\left(1-\frac{1}{3}\omega^2 R^2 \right)~,\\
    A^{NT}_\parallel &= A^{NT}_\perp \cos \theta~,
\end{align}
\end{subequations}
for a rigid spin-zero nucleus \cite{Ru81, HuLv00, Lo54}. Here, $\omega$ is the energy of the incoming and outgoing photon, $M$ is the mass of the nucleus and $R$ is the nuclear charge radius.\\
In contrast to the Thomson scattering \eqref{thomson}, the Rayleigh scattering of light by bound electrons can be understood as the virtual excitation and decay of the target ion and hence is described in the framework of second order perturbation theory \cite{KaKi86}
\begin{widetext}
\begin{equation}
	\label{Msc_equation}
	\begin{aligned}
        A_{\parallel, \perp}^R (\bm k_i, \bm k_f) = & \alpha \int d^3 \bm r_1 \int d^3 \bm r_2 \left( \psi_f^\dag (\bm r_2) \hat{R}^\dag (\bm r_2, \bm k_f, \bm{\epsilon}_{\parallel, \perp}) G(\bm r_2, \bm r_1, E+\omega) \hat{R}(\bm r_1, \bm k_i, \bm{\epsilon}_{\parallel, \perp}) \psi_i(\bm r_1) \right. \\
        & + \left. \psi_f^\dag (\bm r_2) \hat{R}(\bm r_1, \bm k_i, \bm{\epsilon}_{\parallel, \perp}) G(\bm r_2, \bm r_1, E-\omega)  \hat{R}^\dag (\bm r_2, \bm k_f, \bm{\epsilon}_{\parallel, \perp}) \psi_i(\bm r_1) \right).
        \end{aligned}
\end{equation}
\end{widetext}
Here, the initial and final bound-electron wave functions are denoted as $\psi_i (\bm r_1)$ and $\psi_f (\bm r_2)$, describing states with energy $E$ at position $\bm r_{1 ,2}$. Moreover, the intermediate states are described by the  Dirac Coulomb Green's function $G(\bm r_2, \bm r_1, E\pm\omega)$ and $\hat{R}$ is the electron photon interaction operator
\begin{equation}
    \label{ia_operator}
    \hat{R} (\bm r, \bm k, \bm{\epsilon}_{\parallel, \perp}) = \bm{\epsilon}_{\parallel, \perp}\cdot\bm{\alpha} e^{i \bm{k} \bm{r}},
\end{equation}
given here  in Coulomb gauge. In Eq. \eqref{ia_operator}, \(\bm \alpha\) is the vector of Dirac matrices, $ \bm{\epsilon}_{\parallel, \perp}$ is the polarization vector for light linearly polarized within or perpendicular to the scattering plane and $\bm k$ is the wave vector of the photon. Eq. \eqref{Msc_equation} represents the amplitude for Rayleigh scattering by a single bound electron. The wave functions in Eq. \eqref{Msc_equation} are obtained by solving the Dirac equation with a screened potential, taken into account the main part of the electron-electron interactions. The single-electron amplitude can then be used to construct the scattering amplitudes for many-electron (i. e. helium-like) systems, shown by us in Refs. \cite{SuYe18, StYe21}. This can be achieved within the framework of the frozen-core approximation which is well justified for high $Z$ targets and large photon energies \cite{RoKi99, VoYe16}.  For more details about the evaluation of the Rayleigh amplitudes, we refer the reader to Refs. \cite{Rose57, KaKi86, JoCh76, RoKi99, RoSa86, SuYe13, SuIn11, SuYe15}.\\
%
%

\begin{figure}[ht]
    \includegraphics[width=0.99\linewidth]{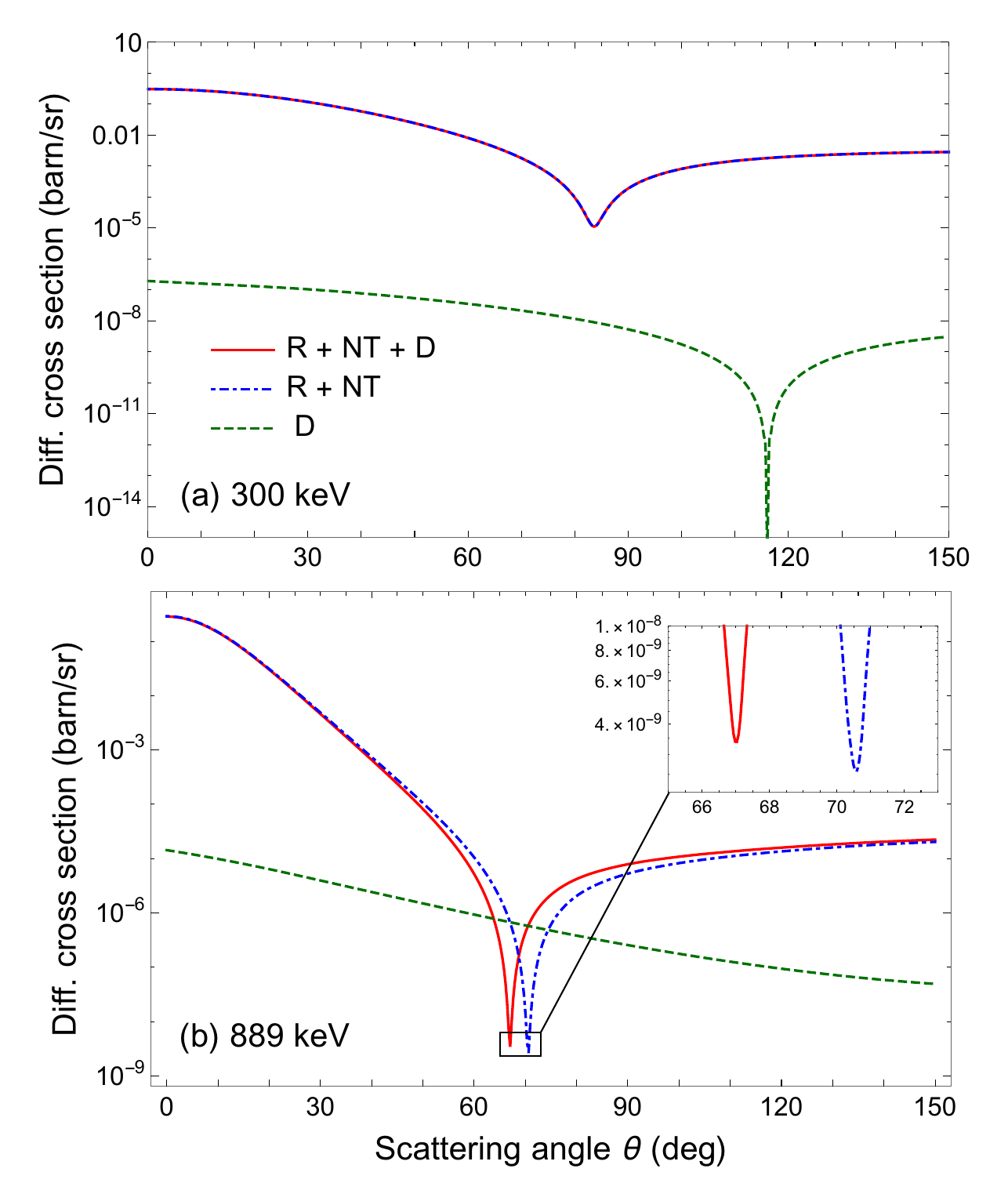}
\caption{Differential cross section for fully linearly polarized light within the scattering plane with photon  energy 300 keV and 889.2 keV, scattered by helium-like tin atoms. The blue dot-dashed curve is Rayleigh and nuclear Thomson scattering (R$+$NT), the green dashed curve depicts Delbrück scattering (D) and the red solid curve shows the interference of the three channels (R$+$NT$+$D).}
\label{cs300}
\end{figure}

\begin{figure}[ht]
    \includegraphics[width=0.99\linewidth]{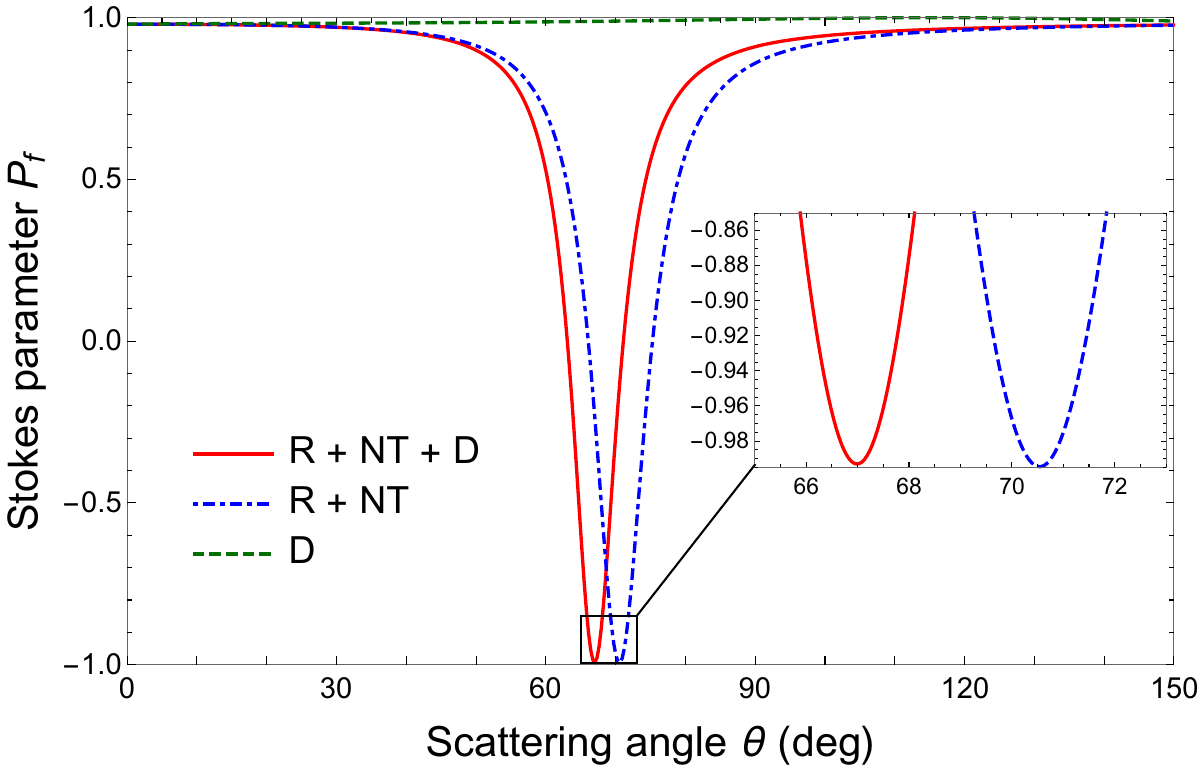}
\caption{Stokes parameters of the scattered light $P_f$ for initial polarization $P_i=0.98$ and photons with energy 889.2 keV scattered by a helium-like tin target. Curves refer to the different scattering channels, similar to Fig. \ref{cs300}.}
\label{p1f}
\end{figure}

The most demanding part of our theoretical analysis is the evaluation of the amplitude for the Delbrück scattering of x-rays by the quantum vacuum. This scattering proceeds via creation and annihilation of a virtual electron positron pair. By taking into account the interaction of the electron (and positron) with the electromagnetic field of the nucleus to all orders in $\alpha Z$, we can write the amplitude for this process as

\begin{equation} \label{MatrixElement}
\begin{aligned}
A_{\parallel, \perp}^D(\bm k_i, &\bm k_f) = \frac{i\alpha}{2\pi} \int_{-\infty}^\infty \text{d}z~\int_{-\infty}^\infty \text{d}z'~\int \text{d}^3\boldsymbol{r}_1~ \\
&\times\int\text{d}^3\boldsymbol{r}_2~\text{Tr}\Big[\hat{R}(\boldsymbol{r}_1,\boldsymbol{k}_i, \bm{\epsilon}_{\parallel, \perp}) G(\boldsymbol{r}_1,\boldsymbol{r}_2,z)\\
&\times \hat{R}^\dagger(\boldsymbol{r}_2,\boldsymbol{k}_f, \bm{\epsilon}_{\parallel, \perp})G(\boldsymbol{r}_2,\boldsymbol{r}_1,z')\Big]\delta (\omega+z-z')~,\\
\end{aligned}
\end{equation}

\noindent where $z$ and $z'$ are the energies of the electron propagators, see~\cite{MILSTEIN1994183}. By expanding the electron-photon interaction operator and the Coulomb Green's functions in Eq.~\eqref{MatrixElement} into angular momentum eigenstates, we can evaluate the angular integrals analytically. The remaining radial integrals are solved with a combination of numerical and analytical methods while the integration over the loop momentum can be performed using a Wick rotation. Further details of this evaluation have been discussed in our recent papers \cite{SoYe22, SoYe23}.\\
By making use of Eqs. \eqref{AmpSum}-\eqref{MatrixElement}, we can evaluate all physical observables of the elastic scattering process. For instance, the angle-differential cross section reads as
\begin{equation}\label{cs}
    \frac{d \sigma}{d \Omega} = \frac{1}{2} \left( |A_{\parallel}|^2 +|A_\perp|^2 \right) +  \frac{1}{2} P_i \left( |A_{\parallel}|^2 - |A_\perp|^2 \right),
\end{equation}
where we assumed that the incident light can be partially linearly polarized and the degree of its polarization is given by Stokes parameter $P_i$. As usual, this parameter is defined as:
\begin{equation}
    P = \frac{I_\parallel-I_\perp}{I_\parallel+I_\perp},
\end{equation}
where \(I_\parallel\) and \(I_\perp\) are the intensities of the light polarized parallel or perpendicular to the scattering plane.\\
Apart from the angular distribution, the scattered light can be also characterized by its degree of linear polarization $P_f$. As shown in Ref. \cite{StYe21}, the latter can be written in terms of its counterpart of the incident x-rays as

\begin{equation}
	\label{P1f_withS}
	P_{f}  =  \frac{|S|^2 (P_{i}+1) - (1-P_{i})}{|S|^2 (P_{i}+1) + (1-P_{i})},
\end{equation}
where we introduced the amplitude ratio \(S(\omega, \theta) \equiv S = A_\parallel/A_\perp\).\\
%
%
%
%
%
\textit{Results.}--- Having briefly recalled the theory of the elastic photon scattering, we are ready now to present the results for both the angle-differential cross section \eqref{cs} and the linear polarization \eqref{P1f_withS} of the scattered x-rays. In our analysis, special attention is paid to the question of whether and how the Delbrück channel affects both observable quantities. For this discussion, we consider the scattering of 300 keV and 889.2 keV photons off a helium-like tin ion. The angle-differential cross section $d \sigma / d \Omega$ calculated for both energies and for completely linearly polarized light, $P_i=1$, is shown in Fig. \ref{cs300}. Here, the green dashed line represents the cross section of Delbrück scattering only (D), the blue dot-dashed line describes the elastic scattering which can proceed via Rayleigh and nuclear Thomson channels (R$+$NT) while finally the red solid line is obtained from the full amplitude \eqref{AmpSum} which takes into account all three channels (R$+$NT$+$D).\\
As follows from our calculations and the upper panel of Fig. \ref{cs300}, the elastic scattering of 300 keV photons by Sn\textsuperscript{48+} is clearly dominated by the Rayleigh contribution. Indeed, $d\sigma^D/d \Omega$ is more than six orders of magnitude smaller than $d\sigma^R/d\Omega$ for almost the entire angular range. The theoretical predictions based on the full amplitude \eqref{AmpSum}, displayed by the red solid line, are almost indistinguishable from the results for $d \sigma^{R} /d \Omega$ and exhibit the standard behavior of the Rayleigh cross section with the minimum around $\theta \approx 83^\circ$ \cite{VoYe16}. This behavior is similar to the $d\sigma^R/d\Omega \sim \cos^2\theta$ \cite{Fl12}, predicted by non-relativistic theory, but is modified by relativistic effects, as discussed in detail in Ref. \cite{VoYe16}. \\
Based on Fig. \ref{cs300}(a)  and the discussion above, we conclude that the contribution of Delbrück scattering can not be ``seen" in experiments with few-hundred keV x-rays as can be performed at synchrotron facilities such as PETRA III. Since the amplitude \eqref{MatrixElement} is known to scale as $A_{\parallel, \perp}^D \sim \omega^2$ \cite{PhysRevD.12.206}, one has to increase the photon energy further to observe the Delbrück channel. In the bottom panel of Fig. \ref{cs300} we display, for example, the differential cross section for the elastic scattering of 889.2 keV photons. Similarly to before, we present here the contributions of Delbrück scattering only, of Rayleigh \textit{plus} nuclear Thomson channels as well as the ``exact" results based on Eq. \eqref{AmpSum}. By comparing upper and lower panels of Fig. \ref{cs300}, we can see that $d\sigma ^{R+NT}/d\Omega \approx d \sigma ^R/ d\Omega$ is only slightly altered for higher photon energies. In particular, the relativistic effects lead to further shift of the minimum to $\theta \approx  70^\circ$ while the absolute values of the cross section for the forward emission angles remain almost the same. In contrast, the behavior of the Delbrück cross section changes significantly. Exhibiting a monotonic descent with emission angle $\theta$, it increases by three orders of magnitude if the energy changes from 300 keV to 889.2 keV and exceeds the R$+$NT cross section in the angular range of $68^\circ \lesssim \theta \lesssim 73^\circ$. One may expect, therefore, that the measurements of the elastic scattering cross section near the minimum of $d\sigma^R/d\Omega$ helps to isolate the Delbrück contribution. For the first time, this idea was proposed by Koga and Hayakawa who investigated the elastic scattering of 1.1 MeV photons by a tin target \cite{KoHa17}. For the analysis of experimental observations, however, it is not sufficient to consider the differential cross section $d\sigma^D/d\Omega$ and $d\sigma^{R+NT}/d\Omega$ separately as proposed e.g. in the aforementioned work. Instead, one has to take into account the interference of all three, D, R and NT channels, see Eqs. \eqref{AmpSum} and \eqref{cs}. As seen from Fig. \ref{cs300}(b), the resulting ``exact" cross section still exhibits the Rayleigh-like behavior, while the position of the minimum is shifted by about $4^\circ$ towards lower scattering angles compared to the R$+$NT case. Exactly this shift is the observable effect of the Delbrück contribution on the angular scattering distribution. \\
Not only the angle-differential cross section but also the linear polarization $P_f$ of the scattered x-rays \eqref{P1f_withS} can be used to explore the Delbrück process. As it was shown in recent works, $P_f$ is very sensitive to the degree of linear polarization of the incident light \cite{SuYe18, StYe21}. In particular, if the incident radiation is slightly depolarized as it usually happens in nowadays synchrotron experiments \cite{BlFr16}, $P_f$ exhibits a narrow dip at the angle of minimum photon emission. Such a behavior is displayed, for example, in Fig. \ref{p1f} where we present the results for the scattering of 889.2 keV photons with initial degree of polarization $P_i = 0.98$. Similar to before, calculations have been performed for Delbrück only and R$+$NT channels, as well as by taking into account all contributions. As seen from the figure, the Delbrück process again leads to the shift in the extremum position of $P_f$ when compared to the R$+$NT predictions. For 889.2 keV photons the shift is again of about $4^\circ$ which can be distinguished with the help of the new generation of Compton polarimeters such as the Compton telescope, presented in Ref. \cite{MiWe22}.\\
%
%
%
\textit{Conclusion.}--- A theoretical investigation was performed for the elastic x-ray scattering off closed-shell atoms and ions. Particular attention was paid to one of the channels of this elastic process --- the Delbrück scattering by the quantum vacuum. Of special interest here is the regime in which the photon energies are below the $e^+e^-$ production threshold. In order to study whether the Delbrück contribution can be ``seen" in the low energy regime, we investigated both the angle-differential cross section and the linear polarization of the scattered photons. Detailed calculations have been performed for the scattering of 300 keV and 889.2 keV x-rays by helium-like tin ions. For the higher energy, it was found that even though Delbrück scattering exceeds the Rayleigh contribution under particular angles, its effect on the experimentally observable angular distribution and the polarization of the outgoing x-rays will be just a shift of their minimum positions by about $4^\circ$. This shift can be observed with the help of modern detection techniques and when comparing with theoretical predictions. This will provide new insights into QED in strong electromagnetic fields. Moreover, this study may serve as benchmark data for studying the giant nuclear resonance channel in the low energy regime where no reliable predictions are published up to our knowledge. 

{\begin{acknowledgments}
This work has been supported by the GSI Helmholtz Centre for Heavy Ion Research under the project BSSURZ1922.
\end{acknowledgments}}
%
%
%
%

\end{document}